\documentclass[pre, amsmath, twocolumn, superspcriptaddress]{revtex4-1}
\usepackage{graphicx}

\begin{document}

\title{Optimizing non-ergodic feedback engines}

\author{Jordan M.~Horowitz}

\author{Juan M.~R.~Parrondo}

\affiliation{Departamento de F\'isica At\'omica, Molecular y Nuclear and GISC, Universidad Complutense de Madrid, 28040 Madrid, Spain}

\date{\today}

\begin{abstract}
{\em Maxwell's demon} is a special case of a feedback controlled system, where information gathered by measurement is utilized by driving a system along a thermodynamic process  that depends on the measurement outcome. The demon illustrates that with feedback one can design an engine that performs work by extracting energy from a single thermal bath. Besides the fundamental questions posed by the demon -- the probabilistic nature of the Second Law, the relationship between entropy and information, etc. -- there are other practical problems related to feedback engines. One of those is the design of optimal engines, protocols that extract the maximum amount of energy given some amount of information. A refinement of the second law to feedback systems establishes  a bound to the extracted energy, a bound that is met by optimal feedback engines. It is also known that optimal engines are characterized by time reversibility. As a consequence, the optimal protocol given a measurement is the one that, run in reverse, prepares the system in the post-measurement state ({\em preparation prescription}). In this paper we review these results and analyze some specific features of the preparation prescription when applied to non-ergodic systems.
\end{abstract}

\maketitle

\section{Introduction}

As pointed out by Maxwell in 1867 with his celebrated {\em demon}, information can be used to extract energy from a single thermal bath~\cite{Leff}. The demon is a special case of feedback control: information about a system is gathered in a measurement, and then the system is driven along a process that depends on that measurement outcome. Subsequent examples by Szilard~\cite{Leff} and others (for example Refs.~\cite{Sagawa2008,Kawai2007,Parrondo2001,Kim2011,Abreu2011}) have revealed that with feedback one can design engines that perform work by extracting energy from a single thermal bath. 

This connection between information and work has been made explicit by a refinement of the second law of thermodynamics in the presence of feedback~\cite{Sagawa2008,Horowitz2010}:  in a thermodynamic process with measurement and feedback, the work $W$ done on a  system is bounded by the difference between the information gained in the measurement $I$ and the change in free energy $\Delta F$ as
\begin{equation}\label{eq:2law}
W\ge \Delta F - kT I
\end{equation}
where $T$ is  temperature and $k$ is Boltzmann's constant. More precisely, $I$ is the {\em mutual information}  Êbetween two random variables: the outcome $m$ of the measurement and the actual value $l$ of the quantity being measured. In an error-free measurement $m=l$, but the concept of mutual information allows us to compute the information gained in a measurement with errors. Mathematically, the mutual information reads
\begin{equation}
I(l;m)=H(l)+H(m)-H(l,m)
\end{equation}
where $H(X)$ is the Shannon entropy of the variable, or set of variables, $X$~\cite{Cover}.
%The mutual information is then the excess of the Shannon entropy of the pair $(l,m)$ if the two variables were independent. 
$I(l;m)=0$ only if $l$ and $m$ are independent, i.e., if the outcome of the measurement is completely uncorrelated with the measured magnitude $l$. On the other hand, if $m=l$ always, $I(l;m)=H(l)=H(m)$ is simply the Shannon entropy of $l$ \cite{Cover}. Notice also that if $z$ is a description of the system finer than $l$ (for instance, the microstate of the system at the instant of measurement), then $I(l;m)=I(z;m)$, provided that the conditional probability of the outcome obeys $\rho(m|l)=\rho(m|z)$.

Besides the fundamental questions posed by the demon -- the probabilistic nature of the Second Law, the relationship between entropy and information, etc.\ -- there are also interesting practical problems related to feedback engines. One of those is how to design optimal engines, i.e., protocols that extract the maximum amount of energy given some amount of information, saturating the bound in Eq.~(\ref{eq:2law})~\cite{Abreu2011,Esposito2011,Kim2011,Kim2011b,Horowitz2011b}. In a sequence of papers, we have shown that these optimal processes  are \emph{reversible}~\cite{Horowitz2011,Horowitz2011b}: indistinguishable from their time-reverse (constructed in a particular manner that will be described later). Building on this intuition, we proposed a method, or a recipe, for designing such optimal feedback processes which we call the \emph{preparation prescription}~\cite{Horowitz2011b}.
Instead of looking for a protocol that extracts all the work, we turn our attention to the time-reversed process and devise a protocol that prepares the post-measurement state.
In this article, we investigate how this method applies to ergodicity-breaking processes, where the phase (or state) space of the system splits into distinct ergodic regions.
The canonical example of this situation is the Szilard engine~\cite{Leff}, where the phase space of a single ideal gas particle confined to a box is divided into two equal halves upon inserting a partition into the center of the box.

The paper is organized as follows. In Sec.~\ref{sec:review}, Êwe briefly review the main results on the energetics of feedback control and the preparation prescription to design optimal engines. In Sec.~\ref{sec:nonergodic}, we analyze the peculiarities of the preparation prescription when applied to non-ergodic systems. In Sec.~\ref{sec:example}, we present an example of optimal design in a multi-particle Szilard engine. Finally in Sec.~\ref{sec:conclusions}, we summarize our results and present our main conclusions.

\section{Reversible Feedback and the Preparation Prescription}\label{sec:review}

We begin with a concise review of the preparation prescription for designing reversible feedback protocols~\cite{Horowitz2011}.
For simplicity, we only consider protocols with one feedback loop.
All of our conclusions can be generalized to the case of a sequence of repeated measurements.

We have in mind a classical system whose position in phase space $\Gamma$ at time $t$ is $z_t$ and is in thermal contact with an ideal thermal reservoir at temperature $T$.
We drive our system away from thermodynamic equilibrium using feedback by varying the system's Hamiltonian (or energy function) $H(z,\lambda)$ through a collection of external parameters $\lambda$. From time $t=0$ to $\tau$, the parameters %$\lambda_t$
 are varied according to a protocol determined by the measurement of a physical observable $M$ at the time $t=t_{\rm meas}$ whose outcomes $m$ occur with conditional probability (or error) $P(m|z_{t _{\rm meas}})$.
The protocol we use, denoted $\Lambda^m=\{\lambda^m_t\}_{t=0}^\tau$, depends on the measurement outcome $m$ only after time $t=t_{\rm meas}$.
During this interval, thermal fluctuations cause the system to follow a random microscopic trajectory $\gamma=\{z_t\}_{t=0}^\tau$. We can define a joint probability distribution ${\mathcal P}[\gamma,\Lambda^m]$ of the trajectory $\gamma$ and the measurement outcome, or equivalently, the implemented protocol $\Lambda^m$. 
The work along this trajectory is $W[\gamma,\Lambda^m$] and the reduction in uncertainty due to the measurement is~\cite{Sagawa2008,Horowitz2010,Horowitz2011}
\begin{equation}\label{eq:I}
i[\gamma,\Lambda^m]=\ln\frac{P(m|z_{t _{\rm meas}})}{p_m}.
\end{equation}
Here, the probability to measure $m$ is $p_m=\int d\gamma\, {\mathcal P}[\gamma,\Lambda^m]$, where $d\gamma$ is a measure on the space of trajectories.
Averaging over all realizations recovers the mutual information $I(z_{t_{\rm meas}};m)=\langle i([\gamma,\Lambda^m]\rangle$ in Eq.~(\ref{eq:2law}).

With every feedback process, we can introduce a related process called the reverse process~\cite{Horowitz2010}, which plays the role of time-reversal in the presence of feedback.
We initiate the reverse process by first randomly selecting a protocol $\Lambda^m$ with probability $p_m$, that is {\em from the distribution of measurement outcomes of the feedback process}. 
Next, we equilibrate the system with the external parameters fixed to $\lambda^m_\tau$, followed by a nonequilibrium driving according to the conjugate reverse protocol $\tilde\Lambda^m=\{\tilde\lambda_{t}\}_{t=0}^\tau$ where $\tilde\lambda_t^m=\lambda_{\tau-t}^m$.
Time-revesal invariance guarantees that each trajectory $\gamma$ of the feedback process has a conjugate twin in the reverse process $\tilde\gamma=\{{\tilde z}_{t}\}_{t=0}^\tau$ where ${\tilde z}_t=z_{\tau-t}^*$ and $*$ denotes momentum reversal, which is observed with probability $\tilde{\mathcal P}[\tilde\gamma, \tilde\Lambda^m]$.

With this setup, we have the result that the distinguishibility of the feedback process measured as the relative entropy, $D(f||g)=\int dx\, f(x)\ln (f(x)/g(x))$, between ${\mathcal P}$ and $\tilde{\mathcal P}$ satisfies \cite{Horowitz2010,Horowitz2011}
\begin{equation}\label{eq:2LawRelEnt}
kT D({\mathcal P}||\tilde{\mathcal P}) = W-\Delta F +kT I \ge0.
\end{equation}
with $I=I(z_{t_{\rm meas}};m)$.
We now see that the optimal thermodynamic process for which $W-\Delta F+kT I=0$ occurs only when $D=0$, which is true if and only if~\cite{Cover}
\begin{equation}\label{eq:reverse}
{\mathcal P}[\gamma,\Lambda^m]=\tilde{\mathcal P}[\tilde\gamma,\tilde\Lambda^m],
\end{equation}
that is only when the feedback process is \emph{indistinguishable} from its reverse~\cite{Horowitz2011}.
This is a microscopic statement of reversibility.
It is consistent with the macroscopic definition, since in a macroscopic reversible process  the same sequence of states also can be traced out  both  forwards and backwards in time.

%With this setup, we can quantify the irreversibility of the process -- the arrow of time -- by measuring how different the forward process is from its reverse.
%We will use an information-theoretic measure of distinguishability, the relative entropy, since it has proven useful.
%It is defined for any pair of functions  $f$ and $g$ as
%\begin{equation}
%D(f||g)=\int dy\, f(y)\ln\frac{f(y)}{g(y)}.
%\end{equation}
%Then we have
%\begin{equation}\label{eq:2LawRelEnt}
%W-\Delta F + I= D({\mathcal P}||\tilde{\mathcal P})\ge0.
%\end{equation}
%We now see that optimal thermodynamic process for which $W-\Delta F+I=0$ occur only when
%\begin{equation}\label{eq:reverse}
%{\mathcal P}[\gamma,\Lambda^m]=\tilde{\mathcal P}[\tilde\gamma,\tilde\Lambda^m],
%\end{equation}
%only when the feedback process is \emph{indistinguishable} from the reverse.
%This is a microscopic statement of reversibility.
%It is consistent with the macroscopic definition, since in a macroscopic reversible process  the same sequence of states also can be traced out  forwards and backwards in time.

Equation~(\ref{eq:reverse}) also offers  insight into how to design an optimal feedback process that extracts the maximum amount of work.
Instead of devising a feedback protocol implemented in response to a particular measurement, we should look for a reversible process.
In particular, let us focus on the evolution at one particular time, immediately after the measurement.
To this end, we integrate Eq.~(\ref{eq:reverse}) over all trajectories passing through $z$ at %a time immediately after 
$t=t_{\rm meas}$, and divide by $p_m$, to deduce the equality of phase space densities conditioned on the protocol (or measurement outcome)
\begin{equation}\label{eq:RevDens}
\rho_m(z,t_{\rm meas})=\tilde\rho_m({\tilde z},\tau-t_{\rm meas}).
\end{equation}
We now see that in a reversible optimal protocol the \emph{post-measurement} state $\rho_m(z,t_{\rm meas})$ -- the state prepared by the measurement -- must be the same as the state \emph{prepared} by the reverse process $\tilde\rho_m({\tilde z},{\tau-t_{\rm meas}})$.  
Our strategy to obtain reversible feedback protocols is then to design a protocol that prepares the post-measurement state~\cite{Horowitz2011b}.
As Eq.~\eqref{eq:RevDens} suggests, by reversing this protocol, we obtain an optimal protocol to implement in the feedback process in response to that measurement outcome.

\section{Preparation in non-ergodic systems}\label{sec:nonergodic}

There is an apparently simple way to reversibly prepare a system in the post-measurement state $\rho_m(z,t_{\rm meas})$ from the initial state of the reverse process, $\tilde\rho(z,0)$: slowly and quasi-statically vary the system Hamiltonian from its initial value $H(z,\tilde\lambda_0^m)$ to $H_m(z)=-kT \ln\rho_m(z,t_{\rm meas})$ so that that post-measurement state is in thermodynamic equilibrium with respect to the new Hamiltonian.
This protocol has been suggested in Refs.~\cite{Hasegawa2010,Takara2010,Esposito2011} and at first sight seems to be the most general procedure for a reversible preparation, since in a reversible process the system must be in equilibrium at any time, in particular, at the beginning and end of the process.

Nevertheless, alternative and more feasible protocols can be devised if the system is not ergodic or if its dynamics presents well separated time scales, as happens in most information processing devices. Consider for instance a system whose phase space $\Gamma$ at the time of measurement, $t_{\rm meas}$, can be decomposed into $n$ distinct ergodic regions $\Gamma_l$  ($\Gamma=\cup_{l=1}^n\Gamma_l$ and $\Gamma_l\cap\Gamma_k=0$ for $l\neq k$). 
This partition of phase space can be the result of a rigorous ergodicity breaking in the system dynamics due to, e.g., barriers higher than the total energy of the system~\cite{Marathe2010} or phase transitions in the thermodynamic limit~\cite{Parrondo2001}. Our analysis also applies to {\em effective ergodicity breaking} resulting when there are slow variables (usually discrete)  whose evolution is goverened, for instance, by jumps over high energy barriers.

\begin{figure*}[htbp]
\begin{center}
\includegraphics[height=5cm]{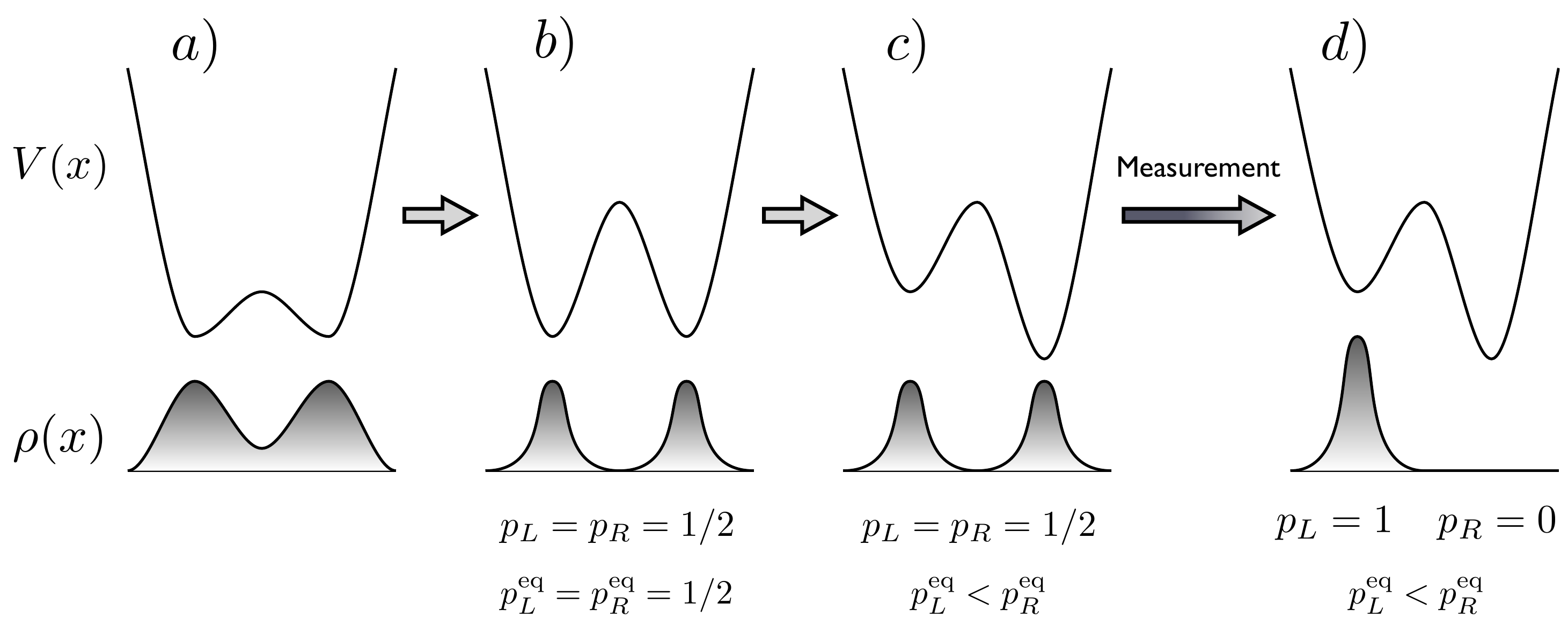}
\caption{An illustration of non-equilibrium states arising from ergodicity breaking and measurement. A  Brownian particle at temperature $T$ moves in a double-well potential $V(x)$ which is modified by an external agent. The potential and the probability density $\rho(x)$ of the position $x$ of the particle are both depicted in the figure. {\em a)} Initially the barrier is low enough for the particle to jump  from one well to the other. {\em b)} The potential barrier is raised up to some value far above $kT$ and an effective ergodicity breaking occurs if we consider a time scale much shorter than the jump rate. The probability that the particle is the left or in the right region is $p_l=1/2$, with $l=L,R$, because ergodicity is broken in a symmetric way. {\em c)} After the transition has occurred, the left well is raised and the right one is lowered. The probability $p_l$ remains 1/2 for $l=L,R$, since jumps do not occur in the time scale of the process, whereas the equilibrium probability,  $p_l^{\rm eq}$ in Eq.~\eqref{eqprob}, changes. {\em d)} After an error-free measurement that finds the particle in the left well, this post-measurement non-equilibrium state is now a probability density with support in the left well, yielding $p_L=1$.}
\label{figpots}
\end{center}
\end{figure*}

We further assume that system is  always locally in equilibrium within each ergodic region and that the measurement is merely the identification of the ergodic region where the system is located. Then, in an error-free measurement the post-measurement state will be  the equilibrium distribution restricted to one of the partitions $\Gamma_l$ at inverse temperature $\beta=1/(kT)$,
\begin{equation}\label{eq:regl}
\rho_{l}(z,t_{\rm meas})=\frac{e^{-\beta H(z,\lambda_{t_{\rm meas}})}}{Z_l} \chi_l(z)
\end{equation}
with $Z_l=\int_{\Gamma_l} e^{-\beta H}$ and $\chi_l(y)$ the characteristic function on $\Gamma_l$ taking the value $1$ when $y\in \Gamma_l$ and $0$ otherwise. 

On the other hand, a measurement with errors can be characterized by the probability  that the actual value of the magnitude is $l$ when the outcome of the measurement is $m$, $p(l|m)$
\footnote{The conditional probability $p(m|l)$ is a more natural way to characterize the error of a measurement device or procedure. To simplify the exposition we  use $p(l|m)$ instead. Both quantities are related by Bayes formula: $p(l|m)=p_lp(m|l)/p_m$. Notice that in feedback control, both $p_l$ and $p_m$ are known (there is no need of a Beayesian {\em  prior}). Feedback uses the information related with thermal fluctuations in a single system, but the statistical properties of such fluctuations are perfectly known.}. 
In this case, when the measurement outcome is $m$, the post-measurement state reads
\begin{equation}\label{eq:postm}
\rho_m(z,t_{\rm meas})=\sum_l p(l|m)\,\frac{e^{-\beta H(z,\lambda_{t_{\rm meas}})}}{Z_l} \chi_l(z)
\end{equation}
According to the preparation prescription, we have to design a protocol that prepares the system in this specific state $\rho_m(z,t_{\rm meas})$. To achieve this goal, it will be illuminating to discuss general features of non-ergodic systems.

In a non-ergodic system equilibration between ergodic regions $\Gamma_l$ is obviously hindered. In a quasi-static process, for instance, the system  is in equilibrium within a region $\Gamma_l$, like the states given by Eqs.~\eqref{eq:regl} and \eqref{eq:postm}, but, for a generic density $\rho(z)$, the probability to be in region $\Gamma_l$ 
\begin{equation}\label{pm}
p_l=\int_{\Gamma_l} dz\,\rho(z)
\end{equation}
will in general differ from its equilibrium value
\begin{equation}
\label{eqprob}
p_l^{\rm eq}=\frac{\int_{\Gamma_l} dx\ e^{-\beta H(x)}}{\int_{\Gamma} dx\, e^{-\beta H(x)}}=\frac{Z_l}{Z}
\end{equation}
In general, the actual $p_l$ depends on the past history and/or the information that we have about the system. For example, if the system becomes non-ergodic by virtue of some symmetry breaking transition, $p_l$ depends on the probability that the system chooses region $l$ {\em at the transition point}. After the transition, the Hamiltonian can change in an arbitrary way, as far as ergodicity is not restored. The equilibrium probability $p_l^{\rm eq}$ in Eq.~\eqref{eqprob} depends on the Hamiltonian at a given time after the transition, whereas $p_l$ depends only on the details of the transition. The probability $p_l$ can also depend on what we know about a system: for instance, after an error-free measurement whose outcome is $l$, $p_l=1$ and $p_k=0$ for all $k\neq l$ [cf.~Eq.~\eqref{eq:regl}]

Fig.~\ref{figpots} presents an illustration that clarifies the meaning of the nonequilibrium probability $p_l$. A  Brownian particle at temperature $T$ moves in a potential $V(x)$, which is modified by an external agent. The potential and the probability density $\rho(x)$ of the position $x$ of the particle are both depicted in the figure. Initially ({\em a}) the barrier is low enough for the particle to jump  from one well to the other.  Then in ({\em b}) the potential barrier is raised up to a value far above $kT$ creating an effective ergodicity breaking for time intervals much smaller than the Kramer's mean time to cross the barrier~\cite{VanKampen}. The probability that the particle is the left or in the right region, $p_l$ with $l=L,R$, is 1/2 because the ergodicity is broken in a symmetric way. After the transition has occurred, one can lower or raise the well in an arbitrary manner as in ({\em c}), as far as the barrier stays far above $kT$. The probability $p_l$ is still 1/2 for $l=L,R$, since jumps do not occur in the time scale of the process. On the other hand, the equilibrium probability  $p_l^{\rm eq}$ in Eq.~\eqref{eqprob}, obviously changes. The state depicted in ({\em c)} is in a nonequilibrium state, although the probability density equilibrates within each well.  Moreover, if we measure (with no error) the position of the particle and find that it is in the left well, the post-measurement nonequilibrium state will be confined in the left well, yielding $p_L=1$ as depicted in ({\em d}). Hence, the nonequilibrium probability $p_l$ depends on the history and also on our knowledge about the state of the system.

Now we can address our main problem: how to prepare a non-ergodic system in the post-measurement state given by Eq.~\eqref{eq:postm}?
Since the state is nonequilibrium, we cannot apply the aforementioned preparation, consisting of a slow transition from the final Hamiltonian $H(z,\tilde\lambda^m_0)$ to $H_m(z)=-kT\ln\rho_m(z,t_{\rm meas})$. 
However, non-ergodicity provides us with a wider range of preparation strategies. The trick is to prepare any other state $\rho_m^\prime(z)$ as long as it reversibly induces the same post-measurement distribution 
\begin{equation}\label{eq:pl}
p(l|m)=\int_{\Gamma_l}dz\,\rho_m^\prime(z)=\int_{\Gamma_l}dz\,\rho_m(z,t_{\rm meas})
\end{equation}
and is in local equilibrium. The key point is that these probabilities $p(l|m)$ depend on the critical point where the ergodicity is broken and not on the final Hamiltonian, as illustrated in Fig.~(\ref{figpots}). Once we prepare a system with the desired probabilities $p(l|m)$, one can adiabatically shift the Hamiltonian towards $H(z,\lambda_{t_{\rm meas}})$ and complete the design of the optimal protocol.

We have applied this method in a previous paper to a multi-particle Szilard engine~\cite{Horowitz2011b}, although we did not carry out an explicit discussion of the role of non-ergodicity. This explicit analysis of the preparation prescription in non-ergodic systems allows us to consider more involved examples, like the one treated in the next section.

\section{Example: Two-Particle Szilard Engine}\label{sec:example}

In this section, we highlight the utility of the preparation prescription for systems with ergodicity breaking using a two-particle Szilard engine.
Previously, Kim {\em et.~al.}~\cite{Kim2011,Kim2011b} investigated the quantum multi-particle Szilard engine using a non-optimal protocol.
In a subsequent article, we then showed how the preparation prescription  could be used to develop an optimal feedback protocol for the classical multi-particle Szilard engine~\cite{Horowitz2011b}.
This section builds on that work to include measurement errors.

\begin{figure}[htb]
\includegraphics[scale=1.1]{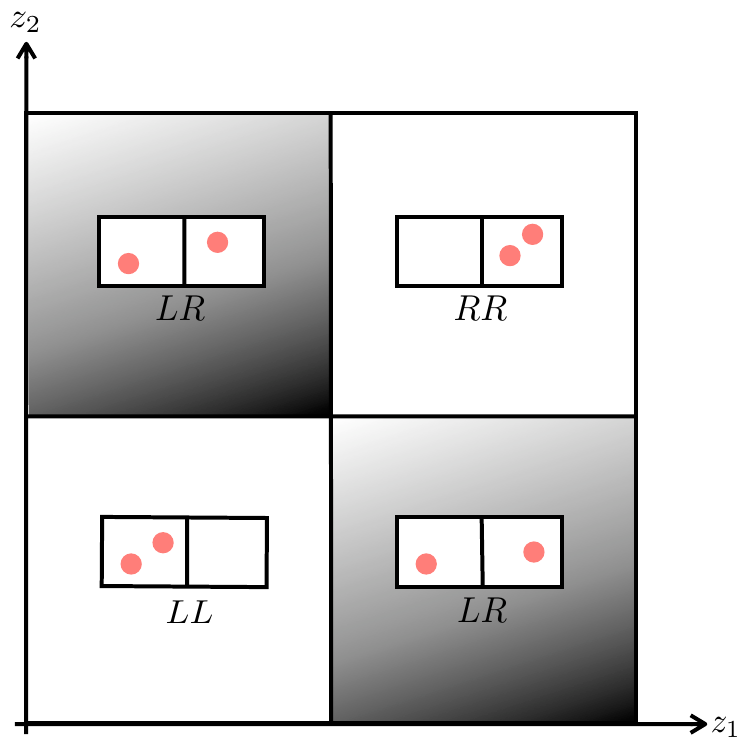}
\caption{Phase space schematic for the two-particle Szilard engine immediately after inserting the partition, with $z_1$ and $z_2$ the phase space positions of the two particles.  Each quadrant corresponds to a ergodic region with a particular arrangement of the two particles: both in the left ($LL$), both in the right ($RR$), or in different halves ($LR$).  The shaded shaded squares highlight the region of phase space where each particle is segregated into a separate half of the box.}
\label{fig:quadrants}
\end{figure}

The two-particle Szilard engine consists of two ideal gas particles confined to a box of volume $V$ connected to a thermal reservoir at temperature $kT=1$.
Furthermore, we take the particles to have a short-ranged, repulsive interaction.
The engine cycle begins with the particles in equilibrium.
We then quickly insert a partition dividing the box into two equal halves, breaking ergodicity.
At that point the phase space of the engine, schematically depicted in Fig.~(\ref{fig:quadrants}), is segregated into three regions that we label $l=\{LL,RR,LR\}$ for two particles in the left half, two in the right, and one in each half.
We then measure $l$ obtaining possible measurement outcomes $m=\{LL,RR,LR\}$.
However, we allow for the possibility that there are errors when both particles are in the same half, but not when they are in separate halves.
Specifically, when $l=LL$ ($RR$) we can mistakenly measure $m=LR$ instead of $LL$ ($RR$) with a probability $\epsilon_{LL}\equiv p(LR|LL)$ [$\epsilon_{RR}\equiv p(LR|RR)$].
Then based on the measurement outcome, we extract work using an optimal, cyclic, isothermal feedback process.

In light of our previous discussion on the preparation prescription (Sec.~\ref{sec:review}), the optimal protocol  will prepare the engine in each ergodic region (or in a distribution over ergodic regions).
%Moreover, since our engine is cyclic, our initial state (of both the feedback and reverse process) will have both particles in equilibrium without the partition.
When both particles are found in the same half of the box the optimal protocol is the same as in the original single-particle Szilard engine.
Namely, we can prepare the engine  with both particles in the left (right) half of the box by inserting the partition along the right (left) wall and then slowly shifting the partition to the center.
Thus, when we find both particles in the same half of the box we can use this protocol, in reverse, to extract the maximum amount of work.

On the other hand, it is more difficult to prepare the engine with each particle in a separate half of the box.
The generic prescription requires that we reversibly prepare the equilibrium distribution for the Hamiltonian $H_m(z)=-kT\ln\rho_m(z,t_{\rm meas})$.
For error-free measurement, this Hamiltonian is infinite in the white quadrants of Fig.~(\ref{fig:quadrants}) and zero in the shaded, which requires infinite interaction energy in disjoint quadrants of phase space.
In a previous article, we demonstrated that using a collection of deep potential wells we could also prepare this scenario, without recourse to such a strange Hamiltonian~\cite{Horowitz2011b}.
In the following, we build on this idea and demonstrate how we can prepare not simply both particles in separate wells, but a distribution over the regions $\{LL,RR,LR\}$ corresponding measurement errors $\epsilon_{LL}$ and $\epsilon_{RR}$ that are rational numbers.

To this end, let us consider the scenario with both particles in the box at equilibrium.
We then slowly lower $K$ potential wells, $n$ in the left half of the box and $K-n$ in the right, to a depth $E<0$ deep compared to the thermal energy $kT=1$, but shallow compared to the interaction energy, so that only one particle can occupy any well at any given time.
This traps each particle in a separate well, occupying a small volume $v$.
Next, we quickly insert the partition, followed by slowly removing the wells. 
As a result, the particles may be confined to either half of the box.
Each particle is in a separate half ($l=LR$) with probability
\begin{equation}
p_{LR}=\frac{2n(K-n)}{K(K-1)}.
\end{equation}
However, this protocol also prepares the system with both particles in the left half ($LL$) and the right half ($RR$) with probabilities
\begin{equation}
p_{LL}=\frac{n(n-1)}{K(K-1)},\qquad p_{RR}=\frac{(K-n)(K-n-1)}{K(K-1)},
\end{equation}
respectively.
As a consequence, it generates a distribution over the different regions, as in Eq.~(\ref{eq:pl}).
Therefore, we can use this protocol (in reverse) as an optimal feedback protocol as long as we use a measurement with an error that results in the distribution $\{p_{LL},p_{RR},p_{LR}\}$ over the regions of phase space given the measurement outcome $m=LR$.
By applying Bayes' theorem, we see this corresponds to measuring $m=LR$ when $l=LL$ with (conditional) probability
\begin{equation}
\epsilon_{LL}\equiv p(LR|LL) = \frac{n-1}{K-n},
\end{equation}
and similarly the error for both particles on the right is
\begin{equation}
\epsilon_{RR}\equiv p(LR|RR) =\frac{K-n-1}{n}.
\end{equation}
For the special case with two wells, one in each half ($K=2$ and $n=1$), we recover error-free measurement ($\epsilon_{LL}=\epsilon_{RR}=0$), which was shown to be optimal in Ref.~\cite{Horowitz2011b}.

%As a result, this protocol prepares the system in the white region of Fig.~(\ref{fig:quadrants}) with probability
%\begin{equation}\label{eq:eps}
%\epsilon=p_{LL}+p_{RR},
%\end{equation}
%even though its designed to prepare the system just in the shaded region.
%, which we can understand in the feedback framework as coming from a measurement errors misidentify both particles in the left (right)half as each particle in separate halves (shaded region) with error probability $\epsilon_{LL}=(n-1)/(K-n)$ [$\epsilon_{RR}=(K-n-1)/n$].

In order to verify that this protocol is in fact optimal for a measurement with errors $\epsilon_{LL}$ and $\epsilon_{RR}$, we now determine the work and information conditioned on measuring the particles in separate halves.
Let us return to our scenario immediately after having inserted the partition and obtained the measurement outcome $m=LR$.
At this point, we lower our $K$ wells very slowly.
As the wells become deeper, the depth approaches a value $E^*\sim kT=1$ at which point ergodicity begins to break, and each particle becomes trapped in a different well.
The exact value of $E^*$ will prove to be inconsequential, but its existence is needed for the calculation.
Since the process is done slowly, the average work done up to that point may be determined as an average over the ratios of the partition functions (the changes in free energy) between the initial state $Z_l$ and the equilibrium state at the moment ergodicity breaks $Z^*_l$ for each $l=\{LL,RR,LR\}$ as
\begin{widetext}
\begin{align}
W_{\rm lower}&=-p_{LL}\ln\frac{Z_{LL}^*}{Z_{LL}}-p_{RR}\ln\frac{Z_{RR}^*}{Z_{RR}}-p_{LR}\ln\frac{Z_{LR}^*}{Z_{LR}} \\
&=-p_{LL}\ln\frac{(1/2)n(n-1) v^{n} e^{-n E^*}}{(1/2)\left(V/2\right)^2}-p_{RR}\ln\frac{(1/2)(K-n)(K-n-1)v^{K-n} e^{-(K-n) E^*}}{(1/2)\left(V/2\right)^2}-p_{LR}\ln\frac{n(K-n) v^K e^{-K E^*}}{\left(V/2\right)^2}.
\end{align}
\end{widetext}
Once the wells have passed $E^*$, each particle is trapped within a separate well, and the work required to lower the wells to the final value $E$ is $w=E-E^*$.
Next, we remove the partition for free.
Then, we begin raising the wells with each particle trapped in a separate well doing a work ${\bar w}=E^*-E$ until we reach $E^*$ again, and the particles begin exploring the entire box.
From this point on, until the wells are completely removed, the work is
\begin{equation}
W_{\rm raise}=-\ln\frac{{\bar Z}}{{\bar Z}^*}=-\ln\frac{V^2/2}{(1/2)K(K-1)v^K e^{-K E^*}}.
\end{equation}
Summing these contributions, we find for the average work conditioned on measuring $m=LR$
\begin{align}
W&=W_{\rm lower}+w+{\bar w}+W_{\rm raise} \\
&=-p_{LL}\ln(4p_{LL}) -p_{RR}\ln(4p_{RR})-p_{LR}\ln(2p_{LR}).
\end{align}
On the other hand, the average information (reduction in uncertainty) can be determined from the formula
\begin{equation}
I=\sum_{l=\{LL,RR,LR\}}p_l\ln \frac{\epsilon_l}{P(LR)}
\end{equation}
by virtue of Eq.~(\ref{eq:I}), where  $P(LR)=1/(2p_{LR})$ is the probability to measure $LR$.
Thus,
\begin{equation}
I=p_{LL}\ln (4p_{LL})+p_{RR}\ln (4p_{RR})+p_{LR}\ln(2p_{LR}),
\end{equation}
and $W+I=0$ as desired.

\section{Conclusions}\label{sec:conclusions}

In this paper we have presented the preparation method as a recipe for designing optimal (or reversible) feedback protocols that extract the maximum amount of energy from a measurement.
In many situations our method reproduces the simplest protocol that exploits the Hamiltonian $H_m(z)=-kT\ln\rho_m(z,t_{\rm meas})$.
However, our method can generate a variety of nontrivial protocols when the system experiences some type of ergodicity breaking.
In our example, the two-particle Szilard engine, we saw that the preparation led to a protocol that exploited a partitioning of phase space and avoided any non-physical Hamiltonians typical of other schemes.

\acknowledgements

This work is funded by  Grants MOSAICO and ENFASIS (Spanish Government),  and MODELICO (Comunidad Autonoma de Madrid).
JMH is supported financially by the National Science Foundation (USA) International Research Fellowship under Grant No.~OISE-1059438.

\bibliographystyle{apsrev4-1.bst} 
\bibliography{Feedback,PhysicsTexts}

%merlin.mbs 2010-03-15 4.21a (PWD, AO, DPC)
\begin{thebibliography}{17}%
\makeatletter
\providecommand \@ifxundefined [1]{%
 \@ifx{#1\undefined}
}%
\providecommand \@ifnum [1]{%
 \ifnum #1\expandafter \@firstoftwo
 \else \expandafter \@secondoftwo
 \fi
}%
\providecommand \@ifx [1]{%
 \ifx #1\expandafter \@firstoftwo
 \else \expandafter \@secondoftwo
 \fi
}%
\providecommand \natexlab [1]{#1}%
\providecommand \enquote  [1]{``#1''}%
\providecommand \bibnamefont  [1]{#1}%
\providecommand \bibfnamefont [1]{#1}%
\providecommand \citenamefont [1]{#1}%
\providecommand \href@noop [0]{\@secondoftwo}%
\providecommand \href [0]{\begingroup \@sanitize@url \@href}%
\providecommand \@href[1]{\@@startlink{#1}\@@href}%
\providecommand \@@href[1]{\endgroup#1\@@endlink}%
\providecommand \@sanitize@url [0]{\catcode `\\12\catcode `\$12\catcode
  `\&12\catcode `\#12\catcode `\^12\catcode `\_12\catcode `\%12\relax}%
\providecommand \@@startlink[1]{}%
\providecommand \@@endlink[0]{}%
\providecommand \url  [0]{\begingroup\@sanitize@url \@url }%
\providecommand \@url [1]{\endgroup\@href {#1}{\urlprefix }}%
\providecommand \urlprefix  [0]{URL }%
\providecommand \Eprint [0]{\href }%
\@ifxundefined \urlstyle {%
  \providecommand \doi  [0]{\begingroup \@sanitize@url \@doi}%
  \providecommand \@doi [1]{\endgroup \@@startlink {\doibase
  #1}doi:\discretionary {}{}{}#1\@@endlink }%
}{%
  \providecommand \doi  [0]{doi:\discretionary{}{}{}\begingroup
  \urlstyle{rm}\Url }%
}%
\providecommand \doibase [0]{http://dx.doi.org/}%
\providecommand \Doi [0]{\begingroup \@sanitize@url \@Doi }%
\providecommand \@Doi  [1]{\endgroup\@@startlink{\doibase#1}\@@Doi}%
\providecommand \@@Doi [1]{#1\@@endlink}%
\providecommand \selectlanguage [0]{\@gobble}%
\providecommand \bibinfo  [0]{\@secondoftwo}%
\providecommand \bibfield  [0]{\@secondoftwo}%
\providecommand \translation [1]{[#1]}%
\providecommand \BibitemOpen [0]{}%
\providecommand \bibitemStop [0]{}%
\providecommand \bibitemNoStop [0]{.\EOS\space}%
\providecommand \EOS [0]{\spacefactor3000\relax}%
\providecommand \BibitemShut  [1]{\csname bibitem#1\endcsname}%
%</preamble>
\bibitem [{\citenamefont {Leff}\ and\ \citenamefont {Rex}(1990)}]{Leff}%
  \BibitemOpen
  \bibinfo {editor} {\bibfnamefont {H.~S.}\ \bibnamefont {Leff}}\ and\ \bibinfo
  {editor} {\bibfnamefont {A.~F.}\ \bibnamefont {Rex}},\ eds.,\ \href@noop {}
  {\emph {\bibinfo {title} {Maxwell's Demon: Entropy, Information,
  Computing}}}\ (\bibinfo  {publisher} {Princeton University Press, New
  Jersey},\ \bibinfo {year} {1990})\BibitemShut {NoStop}%
\bibitem [{\citenamefont {Sagawa}\ and\ \citenamefont
  {Ueda}(2008)}]{Sagawa2008}%
  \BibitemOpen
  \bibfield  {author} {\bibinfo {author} {\bibfnamefont {T.}~\bibnamefont
  {Sagawa}}\ and\ \bibinfo {author} {\bibfnamefont {M.}~\bibnamefont {Ueda}},\
  }\href@noop {} {\bibfield  {journal} {\bibinfo  {journal} {Phys. Rev.
  Lett.},\ }\textbf {\bibinfo {volume} {100}},\ \bibinfo {pages} {080403}
  (\bibinfo {year} {2008})}\BibitemShut {NoStop}%
\bibitem [{\citenamefont {Kawai}\ \emph {et~al.}(2007)\citenamefont {Kawai},
  \citenamefont {Parrondo},\ and\ \citenamefont {Van~den Broeck}}]{Kawai2007}%
  \BibitemOpen
  \bibfield  {author} {\bibinfo {author} {\bibfnamefont {R.}~\bibnamefont
  {Kawai}}, \bibinfo {author} {\bibfnamefont {J.~M.~R.}\ \bibnamefont
  {Parrondo}}, \ and\ \bibinfo {author} {\bibfnamefont {C.}~\bibnamefont
  {Van~den Broeck}},\ }\href@noop {} {\bibfield  {journal} {\bibinfo  {journal}
  {Phys. Rev. Lett.},\ }\textbf {\bibinfo {volume} {98}},\ \bibinfo {pages}
  {080602} (\bibinfo {year} {2007})}\BibitemShut {NoStop}%
\bibitem [{\citenamefont {Parrondo}(2001)}]{Parrondo2001}%
  \BibitemOpen
  \bibfield  {author} {\bibinfo {author} {\bibfnamefont {J.~M.~R.}\
  \bibnamefont {Parrondo}},\ }\href@noop {} {\bibfield  {journal} {\bibinfo
  {journal} {Chaos},\ }\textbf {\bibinfo {volume} {11}},\ \bibinfo {pages}
  {725} (\bibinfo {year} {2001})}\BibitemShut {NoStop}%
\bibitem [{\citenamefont {Kim}\ \emph {et~al.}(2011)\citenamefont {Kim},
  \citenamefont {Sagawa}, \citenamefont {De~Liberato},\ and\ \citenamefont
  {Ueda}}]{Kim2011}%
  \BibitemOpen
  \bibfield  {author} {\bibinfo {author} {\bibfnamefont {S.~W.}\ \bibnamefont
  {Kim}}, \bibinfo {author} {\bibfnamefont {T.}~\bibnamefont {Sagawa}},
  \bibinfo {author} {\bibfnamefont {S.}~\bibnamefont {De~Liberato}}, \ and\
  \bibinfo {author} {\bibfnamefont {M.}~\bibnamefont {Ueda}},\ }\href@noop {}
  {\bibfield  {journal} {\bibinfo  {journal} {Phys. Rev. Lett.},\ }\textbf
  {\bibinfo {volume} {106}},\ \bibinfo {pages} {070401} (\bibinfo {year}
  {2011})}\BibitemShut {NoStop}%
\bibitem [{\citenamefont {Abreu}\ and\ \citenamefont
  {Seifert}(2011)}]{Abreu2011}%
  \BibitemOpen
  \bibfield  {author} {\bibinfo {author} {\bibfnamefont {D.}~\bibnamefont
  {Abreu}}\ and\ \bibinfo {author} {\bibfnamefont {U.}~\bibnamefont
  {Seifert}},\ }\href@noop {} {\bibfield  {journal} {\bibinfo  {journal}
  {Europhys. Lett.},\ }\textbf {\bibinfo {volume} {94}},\ \bibinfo {pages}
  {10001} (\bibinfo {year} {2011})}\BibitemShut {NoStop}%
\bibitem [{\citenamefont {Horowitz}\ and\ \citenamefont
  {Vaikuntanathan}(2010)}]{Horowitz2010}%
  \BibitemOpen
  \bibfield  {author} {\bibinfo {author} {\bibfnamefont {J.~M.}\ \bibnamefont
  {Horowitz}}\ and\ \bibinfo {author} {\bibfnamefont {S.}~\bibnamefont
  {Vaikuntanathan}},\ }\href@noop {} {\bibfield  {journal} {\bibinfo  {journal}
  {Phys. Rev. E},\ }\textbf {\bibinfo {volume} {82}},\ \bibinfo {pages}
  {061120} (\bibinfo {year} {2010})}\BibitemShut {NoStop}%
\bibitem [{\citenamefont {Cover}\ and\ \citenamefont {Thomas}(2006)}]{Cover}%
  \BibitemOpen
  \bibfield  {author} {\bibinfo {author} {\bibfnamefont {T.~M.}\ \bibnamefont
  {Cover}}\ and\ \bibinfo {author} {\bibfnamefont {J.~A.}\ \bibnamefont
  {Thomas}},\ }\href@noop {} {\emph {\bibinfo {title} {Elements of Information
  Theory}}},\ \bibinfo {edition} {2nd}\ ed.\ (\bibinfo  {publisher}
  {Wiley-Interscience},\ \bibinfo {year} {2006})\BibitemShut {NoStop}%
\bibitem [{\citenamefont {Esposito}\ and\ \citenamefont {Van~den
  Broeck}(2011)}]{Esposito2011}%
  \BibitemOpen
  \bibfield  {author} {\bibinfo {author} {\bibfnamefont {M.}~\bibnamefont
  {Esposito}}\ and\ \bibinfo {author} {\bibfnamefont {C.}~\bibnamefont {Van~den
  Broeck}},\ }\href@noop {} {\bibfield  {journal} {\bibinfo  {journal}
  {Europhys. Lett.},\ }\textbf {\bibinfo {volume} {95}},\ \bibinfo {pages}
  {40004} (\bibinfo {year} {2011})}\BibitemShut {NoStop}%
\bibitem [{\citenamefont {Kim}\ and\ \citenamefont {Kim}(2011)}]{Kim2011b}%
  \BibitemOpen
  \bibfield  {author} {\bibinfo {author} {\bibfnamefont {K.~H.}\ \bibnamefont
  {Kim}}\ and\ \bibinfo {author} {\bibfnamefont {S.~W.}\ \bibnamefont {Kim}},\
  }\href@noop {} {\bibfield  {journal} {\bibinfo  {journal} {Phys. Rev. E},\
  }\textbf {\bibinfo {volume} {84}} (\bibinfo {year} {2011})}\BibitemShut
  {NoStop}%
\bibitem [{\citenamefont {Horowitz}\ and\ \citenamefont
  {Parrondo}(2011){\natexlab{a}}}]{Horowitz2011b}%
  \BibitemOpen
  \bibfield  {author} {\bibinfo {author} {\bibfnamefont {J.~M.}\ \bibnamefont
  {Horowitz}}\ and\ \bibinfo {author} {\bibfnamefont {J.~M.~R.}\ \bibnamefont
  {Parrondo}},\ }\href@noop {} {\bibfield  {journal} {\bibinfo  {journal} {New
  J. Phys.},\ }\textbf {\bibinfo {volume} {13}},\ \bibinfo {pages} {123019}
  (\bibinfo {year} {2011}{\natexlab{a}})}\BibitemShut {NoStop}%
\bibitem [{\citenamefont {Horowitz}\ and\ \citenamefont
  {Parrondo}(2011){\natexlab{b}}}]{Horowitz2011}%
  \BibitemOpen
  \bibfield  {author} {\bibinfo {author} {\bibfnamefont {J.~M.}\ \bibnamefont
  {Horowitz}}\ and\ \bibinfo {author} {\bibfnamefont {J.~M.~R.}\ \bibnamefont
  {Parrondo}},\ }\href@noop {} {\bibfield  {journal} {\bibinfo  {journal}
  {Europhys. Lett.},\ }\textbf {\bibinfo {volume} {95}},\ \bibinfo {pages}
  {10005} (\bibinfo {year} {2011}{\natexlab{b}})}\BibitemShut {NoStop}%
\bibitem [{\citenamefont {Hasegawa}\ \emph {et~al.}(2010)\citenamefont
  {Hasegawa}, \citenamefont {Ishikawa}, \citenamefont {Takara},\ and\
  \citenamefont {Driebe}}]{Hasegawa2010}%
  \BibitemOpen
  \bibfield  {author} {\bibinfo {author} {\bibfnamefont {H.-H.}\ \bibnamefont
  {Hasegawa}}, \bibinfo {author} {\bibfnamefont {J.}~\bibnamefont {Ishikawa}},
  \bibinfo {author} {\bibfnamefont {K.}~\bibnamefont {Takara}}, \ and\ \bibinfo
  {author} {\bibfnamefont {D.~J.}\ \bibnamefont {Driebe}},\ }\href@noop {}
  {\bibfield  {journal} {\bibinfo  {journal} {Phys. Lett. A},\ }\textbf
  {\bibinfo {volume} {374}},\ \bibinfo {pages} {1001} (\bibinfo {year}
  {2010})}\BibitemShut {NoStop}%
\bibitem [{\citenamefont {Takara}\ \emph {et~al.}(2010)\citenamefont {Takara},
  \citenamefont {Hasegawa},\ and\ \citenamefont {Driebe}}]{Takara2010}%
  \BibitemOpen
  \bibfield  {author} {\bibinfo {author} {\bibfnamefont {K.}~\bibnamefont
  {Takara}}, \bibinfo {author} {\bibfnamefont {H.-H.}\ \bibnamefont
  {Hasegawa}}, \ and\ \bibinfo {author} {\bibfnamefont {D.~J.}\ \bibnamefont
  {Driebe}},\ }\href@noop {} {\bibfield  {journal} {\bibinfo  {journal} {Phys.
  Lett. A},\ }\textbf {\bibinfo {volume} {375}},\ \bibinfo {pages} {88}
  (\bibinfo {year} {2010})}\BibitemShut {NoStop}%
\bibitem [{\citenamefont {Marathe}\ and\ \citenamefont
  {Parrondo}(2010)}]{Marathe2010}%
  \BibitemOpen
  \bibfield  {author} {\bibinfo {author} {\bibfnamefont {R.}~\bibnamefont
  {Marathe}}\ and\ \bibinfo {author} {\bibfnamefont {J.~M.~R.}\ \bibnamefont
  {Parrondo}},\ }\href@noop {} {\bibfield  {journal} {\bibinfo  {journal}
  {Phys. Rev. Lett.},\ }\textbf {\bibinfo {volume} {104}},\ \bibinfo {pages}
  {245704} (\bibinfo {year} {2010})}\BibitemShut {NoStop}%
\bibitem [{Note1()}]{Note1}%
  \BibitemOpen
  \bibinfo {note} {The conditional probability $p(m|l)$ is a more natural way
  to characterize the error of a measurement device or procedure. To simplify
  the exposition we use $p(l|m)$ instead. Both quantities are related by Bayes
  formula: $p(l|m)=p_lp(m|l)/p_m$. Notice that in feedback control, both $p_l$
  and $p_m$ are known (there is no need of a Beayesian {\protect \em prior}).
  Feedback uses the information related with thermal fluctuations in a single
  system, but the statistical properties of such fluctuations are perfectly
  known.}\BibitemShut {Stop}%
\bibitem [{\citenamefont {Van~Kampen}(2007)}]{VanKampen}%
  \BibitemOpen
  \bibfield  {author} {\bibinfo {author} {\bibfnamefont {N.~G.}\ \bibnamefont
  {Van~Kampen}},\ }\href@noop {} {\emph {\bibinfo {title} {Stochastic Processes
  in Physics and Chemistry}}},\ \bibinfo {edition} {3rd}\ ed.\ (\bibinfo
  {publisher} {Elsevier Ltd., New York},\ \bibinfo {year} {2007})\BibitemShut
  {NoStop}%
\end{thebibliography}%

\end{document}